\documentclass[final,1p,times]{elsarticle}

\usepackage{amssymb}
\usepackage{lipsum}
\usepackage{amsthm}

\usepackage{amsfonts}
\usepackage{mathrsfs}
\usepackage{systeme, mathtools}

\usepackage{tikz-cd}
\usepackage{xcolor}
\usepackage{graphicx}
\usepackage{dcolumn}
\usepackage{bm}
\usepackage{appendix}
\usepackage{url}
\usepackage{hyperref}
\hypersetup{hidelinks}
\numberwithin{equation}{section}
\newtheorem{theorem}{Theorem}[section]

\newtheorem{proposition}[theorem]{Proposition}

\theoremstyle{definition}

\newtheorem{definition}[theorem]{Definition}

\newcommand{\R}{\mathbb{R}}
\newcommand{\C}{\mathbb{C}}
\newcommand{\bra}{\langle}
\newcommand{\ket}{\rangle}
\newcommand{\Dom}[1]{\mathrm{Dom}\left(#1\right)}

\journal{Nuclear Physics B}
\makeatletter
\def\ps@pprintTitle{%
 \let\@oddhead\@empty
 \let\@evenhead\@empty
 \let\@oddfoot\@empty
 \let\@evenfoot\@empty}
\makeatother

\begin{document}

\begin{frontmatter}

\title{Unitary equivalence in Generalized Uncertainty Principle theories}

\author[Sap]{Sebastiano Segreto}
\ead{sebastiano.segreto@uniroma1.it}
\author[Sap,INFN]{Matteo Bruno}
\ead{matteo.bruno@uniroma1.it}

\affiliation[Sap]{organization={Physics Department, Sapienza University of Rome},
            addressline={P.za Aldo Moro 5}, 
            city={Rome},
            postcode={00185}, 
            state={},
            country={Italy}}
\affiliation[INFN]{organization={INFN, Sezione               di Roma 1},
            addressline={P.le Aldo Moro 2}, 
            city={Rome},
            postcode={00185}, 
            state={},
            country={Italy}}

\begin{abstract}
We analyze the issue of unitary equivalence within Generalized Uncertainty Principle (GUP) theories in the one-dimensional case. For a deformed Heisenberg algebra, its representation in terms of Hilbert space and conjugate operators is not uniquely determined, raising the question of whether different realizations of the same algebra are equivalent and describe the same physics. After proposing a definition of a quantum GUP theory, we establish conditions for unitary equivalence. Using this framework, we rigorously prove that two commonly used representations are unitarily equivalent, specifying the conditions under which this equivalence holds. We demonstrate this equivalence explicitly by providing a unitary map and showing how both GUP formulations yield the same physical results in two examples: the quantum harmonic oscillator and a free-falling particle. Finally, we discuss a case in which equivalence fails, suggesting that a generalization of the Stone-von Neumann theorem may not be possible within the GUP framework under our definition of unitary equivalence.
\end{abstract}

\begin{keyword}
generalized uncertainty principle \sep Stone-von Neumann theorem 



\end{keyword}

\end{frontmatter}

 \section{Introduction}\label{Xsec1-1}
  Generalized uncertainty principle (GUP) theories are widely recognized for providing an effective framework, originally rooted in String theory considerations \cite{Polchinski:2014mva, Amati:1987wq, Amati:1988tn, Gross:1987ar, Gross:1987kza, Konishi:1989wk}, that describes an alternative structure of space, consistent with what a quantum theory of gravity is expected to yield. These theories are quantum, non-relativistic models based on a deformation of the usual Heisenberg algebra between conjugate quantum operators \cite{kempf1994uncertainty,Kempf:1994su,Maggiore:1993kv,Maggiore:1993rv,ali2009discreteness,Bosso:2023aht,Segreto:2022clx,Segreto:2024vtu}.
  The new quantum picture that emerges from this fundamental modification is drastically different from the ordinary one. Specifically, the changes introduced in the algebra can manifest in two main features: on the one hand, the emergence of a non-zero minimal uncertainty in position, which can be interpreted as the appearance of a fundamental minimal length or, more generally, a minimal structure in configuration space; on the other hand, in higher-dimensional scenarios, a non-commutativity between position or coordinate operators arises, which can be viewed as the onset of a non-commutative geometry in the configuration space of the theory. As we anticipated, these novel characteristics, directly embedded in the quantum formalism, result in a structure of space that closely resembles what one might expect, on general grounds, from a fundamentally quantum nature of spacetime. In particular, the usual notions of position and localization are profoundly altered, necessitating new concepts and tools to properly describe a system's behavior in configuration space and, more broadly, its dynamics \cite{Segreto:2024vtu}. \\
  The GUP framework boasts a very wide range of applications in various physical contexts, each of which may require a different conceptual frame for its formulation \cite{tawfik2014generalized,Bosso:2023fnb,bosso2021generalized,amelino2006black,Battisti:2008rv,Ong:2018zqn,segreto2025dynamics,bruno2024generalized}. Indeed, while the GUP can serve as an effective means of introducing corrections to the dynamics of a system at scales where quantum gravity effects become significant, i.e. the Planck scale, it can also be interpreted, when used as a quantization scheme for gravitational systems, as a more faithful method of quantizing gravitational degrees of freedom. Regardless of the conceptual standpoint, the literature offers numerous applications with compelling results, demonstrating both the versatility of the framework and its potential to profoundly affect a system's dynamics. \\
  Nevertheless, the reliability of these results, at least from a formal perspective, fundamentally depends on the correct construction of the mathematical structures defining the theories. Modifying the algebra of fundamental quantum operators must be approached with care to coherently construct the new quantum theory, even in its more formal aspects---areas that are often overlooked in recent discussions of the subject. \\
  In this paper, we address a very general issue concerning the construction of a GUP theory: the unitary equivalence among different formulations of the same GUP theory arising from a given modified algebra. This topic is, of course, of broad importance in physics. In standard quantum theory, it is well established that different, yet unitarily equivalent, representations of the Heisenberg algebra are possible, as guaranteed by the celebrated Stone-von Neumann theorem \cite{v1931eindeutigkeit, neumann1932ueber, stone1930linear, stone1932one, rosenberg2004selective, summers2001stone}. Unitary equivalence places all such formulations on equal footing, as they necessarily describe the same physics. In light of this, it is entirely natural to pose the same question in the broader context of the GUP framework. This issue is not only relevant at a formal level, but also aids in a deeper understanding of the deformed canonical structure of these models. This could be particularly relevant in constructing deformed quantum canonical transformations, which could reveal themselves to be valuable in solving the dynamics of a given system---a task often complicated by the more intricate GUP representations of the operators involved.\\
  We will limit our analysis and considerations to the one-dimensional scenario. \\
  In general, given a modified algebra, different Hilbert spaces can be chosen for its representation, that is, for the action of the fundamental operators. A priori, it is not obvious that two different formulations derived from the same algebra are physically equivalent. To determine under what conditions equivalence holds, we must first agree on the definition of a GUP theory, identify the main objects of interest and establish some minimal properties that the conjugate operators and the deformation operator-valued function $f$, along with their domains, must satisfy. This procedure allows us to identify a GUP formulation with the Hilbert space $\mathcal{H}$ chosen for its representation and the quantum conjugate operators $q, p$, along with their specific representation, defined on some proper domains where they turn out to be self-adjoint. These objects define a triple $(\mathcal{H}, q, p)$ by which a GUP theory can be uniquely labeled. Given two such triples, if a unitary transformation exists between the two Hilbert spaces---subject to certain precise requirements---then they represent the same one-dimensional GUP theory and thus describe the same physics. Clearly, if such a transformation does not exist, we must conclude that we are dealing with two distinct physical theories, each with different physical implications. \\
  Our analysis mainly focuses on demonstrating the unitary equivalence between two major GUP representations found in the literature. On one side, there is the GUP representation introduced by Kempf, Mangano and Mann in \cite{Kempf:1994su}, where the Hilbert space employs a modified Lebesgue measure to ensure the symmetry of the position operator, thereby determining its momentum representation; on the other, there is a commonly used representation where the Hilbert space is the standard one and the usual operators $Q$ and $P$, satisfying the ordinary Heisenberg algebra, are introduced as auxiliary operators. We discuss in detail---most of which is developed in the appendix---the functional structure of these two representations, highlighting crucial aspects regarding the functional properties of the operators involved, and then proceed to show that, only under specific conditions, these two formulations are indeed unitarily equivalent. \\
 To explicitly illustrate how the entire machinery operates, we analyze, within both settings, two very simple yet instructive examples: the quantum harmonic oscillator and the free fall of a quantum particle. While the first problem is addressed by fixing a specific algebra, the second is treated for any generic algebra. By solving the problems in both formulations and using the previously constructed unitary transformations, we demonstrate how, separately for each of them, the dynamics map onto each other, revealing their complete physical equivalence in both scenarios. \\
 Finally, before presenting our conclusions, we briefly comment on an important counterexample of non-unitary equivalence, involving the previously discussed GUP formulations and the so-called polymer-like formulations, which are sometimes employed in certain physical contexts \cite{Barca:2021epy}. These GUP representations are constructed using a representation reminiscent of that used in another non-standard quantum theory, namely Polymer Quantum Mechanics (PQM), which is inspired by Loop Quantum Cosmology (LQC) \cite{corichi2007hamiltonian, corichi2007polymer}. Utilizing the tools developed in this work, we show that, at least for those GUP theories that lead to the emergence of a minimal length, these two formulations are not unitarily equivalent, that is, they describe different physics at the quantum level. This provides significant evidence that a generalization of the Stone-von Neumann theorem may not exist in this context, at least according to our definitions of unitary equivalence.
 
\section{Equivalence and Consequences}
\begin{definition} \label{def:gup-th}
    A Hilbert space associated with a $1$-dimensional Generalized Uncertainty Principle theory is a Hilbert space $\mathcal{H}$ together with two distinguished densely-defined linear operators $q,p$ and the data of a measurable real positive function $f$ on the spectrum $\sigma(p)$, continuous and differentiable almost everywhere, such that:
    \begin{enumerate}
        \item $q$ and $p$ are self-adjoint;
        \item the commutator $[q,p]$ has a dense domain and satisfies $\Dom{[q,p]}\subset\Dom{f(p)}$. Moreover, on $\Dom{[q,p]}$ \[[q,p]=i\hbar f(p).\]
    \end{enumerate}
    Here, $f(p)$ is interpreted in the sense of functional calculus.
\label{Xenun2-1}
\end{definition}
The first condition is linked to the physicality of $q$ and $p$, as position and momentum. The second one characterizes the GUP theory via the function $f$. In this work, we will consider $f$ continuous and bounded below by a positive number. The function $f$ characterizes the GUP theory in the following sense:
\begin{definition}\label{def:same_th}
    We say that two triples $(\mathcal{H}_1,q_1,p_1)$ and $(\mathcal{H}_2,q_2,p_2)$ are Hilbert spaces associated with the same $1$-dimensional Generalized Uncertainty Principle theory if they satisfy the previous definition with a function $f_1$ and $f_2$, respectively, and $\sigma(p_1)=\sigma(p_2)$ and $f_1=f_2$ almost everywhere on the spectrum.
\label{Xenun3-2}
\end{definition}

\begin{definition}\label{def:equiv}
    Two Hilbert spaces $(\mathcal{H}_1,q_1,p_1)$ and $(\mathcal{H}_2,q_2,p_2)$ associated with the same $1$-dimensional Generalized Uncertainty Principle theory are equivalent if there exists a unitary transformation $U:\mathcal{H}_1\to\mathcal{H}_2$ such that \[Uq_1U^{-1}=q_2,\ \, and\ \,Up_1U^{-1}=p_2.\]
\label{Xenun4-3}
\end{definition}
\subsection{Impact on usual Hilbert spaces}\label{Xsec3-2.1}
These definitions have an immediate impact on transformations between Hilbert spaces, introducing a sort of ``quantum canonical transformation'' within the GUP theories context.\\
In this section, we will discuss two of the most relevant and used functional settings for a GUP theory, examining their unitary, thus physical, equivalence. \\
The first most common Hilbert space used in the GUP framework is represented by $\mathcal{H}_1:= L^2\left(\R,\frac{du}{f(u)}\right)$. Here, the two distinguished operators are defined as:
\begin{align}
\label{first_triple}
    \begin{split}
    &(p\psi)(u)=u\psi(u),\\
    &(q\psi)(u)=i\hbar f(u)\frac{d}{du}\psi(u).
    \end{split}
\end{align}

Alongside this setting, another commonly employed space is the usual $\mathcal{H}_2:= L^2(\R,dv)$. On such a space, the standard commuting self-adjoint operators $Q,P$ are usually introduced and defined as:
\begin{align}
\begin{split}
    &(P\psi)(v)=v\psi(v)\\
    &(Q\psi)(v)=i\hbar \frac{d}{dv}\psi(v)
\end{split}
\label{Xeqn2-2.2}
\end{align}
This space can be associated with a $1$-dimensional GUP theory considering two operators $q(Q,P)$ and $p(Q,P)$ as functions of $Q,P$ and requiring 1 and 2 of Def.~\ref{def:gup-th}. The pair $Q,P$ has to be considered solely as self-adjoint auxiliary operators, without any physical meaning, introduced to exploit a suitable representation for the physical operators $q,p$.
The imposition $[q(Q,P),p(Q,P)]=i\hbar f(p)$ gives us a series of constraints on the functions. Choosing $p=P$, $q$ must be of the form
\begin{align*}
    q_a(Q,P)=af(P)Q+(1-a)Qf(P),
\end{align*}
where $a$ is a real parameter between 0 and 1. \\
In both settings, the two fundamental operators $q,p$ must be defined along with their domains. Indeed, in general, operators are linear maps together with the domain on which they act (usually a dense subset of the whole Hilbert space). In particular, considerations about domains become extremely important when discussing unbounded operators, which is exactly the case of the fundamental operators of quantum mechanics. The domain of the operators $q,p$ have to be identified to fulfill the two requests of Def.~\ref{def:gup-th}, in particular the self-adjointness demand.\\
The momentum operator $p$ is easily seen to be essentially self-adjoint in both our functional frameworks and its unique self-adjoint extension---i.e. the unique domain on which it turns out to be self-adjoint---is represented by the following domains:
\begin{itemize}
    \item the operator $p$ defined in \eqref{first_triple} is self-adjoint with the domain
    \begin{align}
        \Dom{p}:=  \left\{\psi \in \mathcal{H}_1 \, \bigg| \, \int_{\mathbb{R}} \frac{du}{f(u)} | u \, \psi(u)|^2 < + \infty\right \}\subset L^2\left(\R,\frac{du}{f(u)}\right), \label{mom_dom_1}
    \end{align}
    \item the operator $p$ defined in \eqref{second_triple} is self-adjoint with the domain
    \begin{align}
        \Dom{p}:=  \left\{\psi \in \mathcal{H}_2 \, \bigg| \, \int_{\mathbb{R}} dv\, | v\,\psi(v)|^2 < + \infty\right\}\subset L^2(\R,dv). \label{mom_dom_2}
    \end{align}
\end{itemize}
On the other hand, the analysis of the position operator $q$ is less trivial (see e.g. \cite{Kempf:1994su, Segreto:2022clx, Segreto:2024vtu} for a comprehensive discussion).\\
The first thing to correctly address is the symmetry of the operator.
In both settings, we can consider the domain of definition to be the space $C^{\infty}_c(\mathbb{R})$ of all smooth, compactly supported functions, seen as a dense subspace in the respective Hilbert spaces.
On these domains, in $\mathcal{H}_1$, the symmetry is ensured by the purposely introduced deformed Lebesgue measure $\frac{du}{f(u)}$, while in $\mathcal{H}_2$ symmetry is simply achieved with the choice of $a=\frac{1}{2}$.
In this last setting, the two distinguished operators can be written as:
\begin{align}
\label{second_triple}
    \begin{split}
    &q(Q,P)=\frac{1}{2}\left(f(P)Q+Qf(P)\right),\\
    &p(Q,P)=P.
\end{split}
\end{align}

In general, for the $q$ operator, essential self-adjointness is not guaranteed; rather, the operator admits infinitely many self-adjoint extensions \cite{moretti2013spectral}.
This technical fact has profound consequences on the physical level, since the loss of essential self-adjointness induces the emergence of minimal structure in the configuration space of the theory, such as a minimal length.
Compactly, the domains on which the operator turns out to be self-adjoint can be written as:
\begin{align} \label{sa_ext_q}
    \Dom{q_{\lambda}}:= \Dom{\,\overline{q}\,}+(\mathbb{I}-U_\lambda)(\mathrm{Ker}(q^{\dagger}-\mathrm{i}\mathbb{I})),
\end{align}
where $\overline{q}$ is the closure of the symmetric operator $q$ defined as above and $U_\lambda$ is a family of unitary operators which are in one-to-one correspondence with all the possible self-adjoint extensions of $q$, parametrized by $\lambda$. The existence and the form of the unitary operators $U_{\lambda}$ depend on the behavior of the function $f$ that characterizes the GUP theory. Note that, whenever $U_\lambda=\mathbb{I}$ for all $\lambda$, we recover the case in which $q$ is essentially self-adjoint. In particular the following chain of equalities holds $q_{\lambda}=\overline{q}=q^{\dagger \dagger}=q^{\dagger}$ \cite{moretti2013spectral}. \\
By means of these domains, as we explicitly show in \ref{App:comm}, it is straightforward to see that conditions (2) of Def.~\ref{def:gup-th} is satisfied for any fixed extension, that is \[\Dom{[q_\lambda,p]} \underset{\mathcal{H}_{1,2}}{\subset} \Dom{f(p)}.\]
Hence, we will call $\mathcal{T}^\lambda$ the triple given by the space $L^2\left(\R,\frac{du}{f(u)}\right)$, the unique self-adjoint extension of the operator $p$ and the self-adjoint extensions of $q$ indexed by $\lambda$, with $q,p$ defined as in \eqref{first_triple}. While $\mathcal{K}^\lambda$ is the triple given by the space $L^2(\R,dv)$, the unique self-adjoint extension of the operator $p(Q,P)$ and the self-adjoint extensions of $q(Q,P)$ indexed by $\lambda$, with $Q,P$ defined as in \eqref{second_triple}. For the reasons discussed above, both $\mathcal{T}^{\lambda}$ and $\mathcal{K}^{\lambda}$ are Hilbert spaces associated with a $1$-dimensional GUP theory.
\begin{proposition} \label{core_props}
    The two Hilbert spaces $\mathcal{T}^{\lambda}$ and $\mathcal{K}^{\lambda}$ associated with the same $1$-dimensional GUP theory are equivalent for any fixed $\lambda$.
\label{Xenun1-1}
\end{proposition}
\begin{proof}
    The first step of the proof is the construction of the unitary map and then, the verification of the proper behavior of the distinguished operators.\\
    The unitary map is
    \begin{align} \label{u-map}
        \begin{matrix}
            U: & L^2\left(\R,\frac{du}{f(u)}\right) & \to & L^2\left(\R,dv\right);\\
             & \psi(u) & \mapsto & \tfrac{1}{\sqrt{f(v)}}\psi(v).
        \end{matrix}
    \end{align}
    We can easily check that it is an isometry
    \begin{align}
        \bra U\psi|U\psi\ket_{\mathcal{H}_2}=\int_{\R} \overline{(U\psi)(v)}(U\psi)(v) dv=\int_{\R} \overline{\psi(u)}\psi(u) \tfrac{du}{f(u)}=\bra\psi|\psi\ket_{\mathcal{H}_1},
    \label{Xeqn8-2.8}
\end{align}
    with inverse
    \begin{align}
        \begin{matrix}
            U^{-1}: & L^2\left(\R,dv\right) & \to & L^2\left(\R,\frac{du}{f(u)}\right);\\
             & \phi(v) & \mapsto & \sqrt{f(u)}\phi(u).
        \end{matrix}
    \label{Xeqn9-2.9}
\end{align}
    Let us call $p_1$ the operator $p$ on $ L^2\left(\R,\frac{du}{f(u)}\right)$ and $p_2$ the operator $p(Q,P)$ on $ L^2\left(\R,dv\right)$, and the same for the operators $q$. \\ First, we prove that $Up_1U^{-1}=p_2$:
    \begin{align}
        \begin{split}
            (Up_1U^{-1}\phi)(v)&=Up_1\left(\sqrt{f(u)}\phi(u)\right)=U\left(u\sqrt{f(u)}\phi(u)\right)\\
        &=\tfrac{1}{\sqrt{f(v)}}v\sqrt{f(v)}\phi(v)=v\phi(v).
        \end{split}
    \label{Xeqn10-2.10}
\end{align}
    Now, we prove that $Uq_1U^{-1}=q_2$:
    \begin{align}
        \begin{split}
            (Uq_1U^{-1}\phi)(v)&=Uq_1\left(\sqrt{f(u)}\phi(u)\right)\\
        &=U\left(i\hbar f(u)\left(\frac{1}{2\sqrt{f(u)}}\frac{df}{du}\phi(u)+\sqrt{f(u)}\frac{d\phi}{du}\right)\right)\\
        &=i\hbar \frac{1}{2}\frac{df}{dv}\phi(v)+i\hbar\frac{d\phi}{dv}=\frac{i\hbar}{2}\left(\frac{d}{dv}(f\phi)+f\frac{d\phi}{dv}\right)
        \end{split}
    \label{Xeqn11-2.11}
\end{align}

    This shows how the unitary transformation correctly maps the action of the operators in the two Hilbert spaces. To proceed, we need to prove that the map $U$ maps the operators' domains properly.\\
    Before doing that, we provide a consistency check verifying that the unitary maps preserve the non-extended operators' functional properties, namely that it maps bijectively the kernel of the operator $\left(q_1^{\dagger} \pm i \mathbb{I}_{\mathcal{H}_1}\right)$ into $\left(q_2^{\dagger} \pm i \mathbb{I}_{\mathcal{H}_2}\right)$, and the same for $p_1$, $p_2$. \\
    For the momentum operators $p_1, p_2$, acting as multiplicative operators, the kernels are trivial and correctly mapped into each other by the unitary transformation $U$.
    This means that the essentially self-adjoint operator $p_1$ is mapped into the essentially self-adjoint operator $p_2$ and vice versa. \\
    For the position operators $q_1, q_2$, the dimension of the kernels is determined by the solutions of some differential equations \cite{moretti2013spectral}. \\
    For the operator $q_1$, the differential equation is:
    \begin{align}
        i \hbar f(u) \partial_{u} \psi(u) \pm i \psi(u)=0,
    \label{Xeqn12-2.12}
\end{align}
    where the derivative has to be intended as a weak derivative.
    The solutions are:
    \begin{align} \label{sol_def_ind_eqs}
        \psi_{\pm}(u)=k \exp\left(\mp\frac{1}{\hbar} \int_{0}^{u}\frac{1}{f(s)}ds\right), \quad \text{a.e. in }\mathbb{R}, \, k \in \mathbb{C}.
    \end{align}
    For $q_2$, the differential equation is:
    \begin{align}
        i \hbar f(v) \partial_{v} \phi(v) + i \frac{\hbar}{2} \phi(v) \partial_{v}f(v) \pm i \phi(v)=0,
    \label{Xeqn14-2.14}
\end{align}
    whose solutions are:
    \begin{align}
        \phi_{\pm}(v)=k \frac{1}{\sqrt{f(v)}}\exp\left(\mp \frac{1}{\hbar} \int_{0}^{v}\frac{1}{ f(t)}dt\right).
    \label{Xeqn15-2.15}
\end{align}
    Whether or not these solutions belong to the domains of $q^{\dagger}_1$ and $q^{\dagger}_2$, respectively, determines the dimensions of the kernels of interest. \\
    As it is straightforward to realize, whenever these solutions lie in the respective Hilbert spaces, they are mapped into each other by the transformations $U$ and $U^{-1}$, which in turn implies that the kernels of the operators $\left(q_i^{\dagger} \pm i \mathbb{I}_{\mathcal{H}_i}\right)$ in the two Hilbert spaces are unitarily mapped into each other as well.
    This proves that the defined operators $q_1, q_2$ correctly share the same functional properties with respect to self-adjointness. \\
    Finally, to complete the proof, we need to show that the map $U$ and its inverse correctly map the self-adjointness domains of the involved operators.
    First, we consider the domains of the momentum operators in $\mathcal{H}_1$ and $\mathcal{H}_2$.
    Consider an element $\psi(u) \in \Dom{p_1}$. The unitary map $U$ sends this element to the element $\psi(v)/\sqrt{f(v)}$, for which it is true that:
    \begin{align}
        ||p_2U\psi||^2_{\mathcal{H}_2}=\int_{\mathbb{R}} dv \left|v^2 (U\psi)(v)\right|^2=\int_{\mathbb{R}}\frac{dv}{f(v)}|v \psi(v)|^2
     < + \infty.    \label{Xeqn16-2.16}
\end{align}
   From the condition $\psi(u) \in \Dom{p_1}$, stated in \eqref{mom_dom_1}, the integral is finite. Hence this proves that $U(\Dom{p_1})\subset  \Dom{p_2}$.
   Consider now an element $\phi(v) \in \Dom{p_2}$, the inverse unitary transformation $U^{-1}$ maps this element to the element $\sqrt{f(u)}\phi(u)$, for which the following holds:
   \begin{align}
       ||p_1U^{-1}\phi||^2_{\mathcal{H}_1}=\int_{\mathbb{R}}\frac{du}{f(u)}\left|\sqrt{f(u)}\,u \,\phi(u)\right|^2=\int_{\mathbb{R}}|u \phi(u)|^2 < +\infty,   \label{Xeqn17-2.17}
\end{align}
    by the hypothesis $\phi(v) \in \Dom{p_2}$, whose implications are stated in \eqref{mom_dom_2}.
    This proves that $U^{-1}(\Dom{p_2})\subset\Dom{p_1}$ and allows us to conclude that, correctly, $U( \Dom{p_1})=\Dom{p_2}$.

    We can proceed with the domains of the position operators.
    With reference to the domain formula \eqref{sa_ext_q}, to prove that the unitary transformation along with its inverse maps correctly the domains involved, we can limit ourselves to the domain of the closure of the operators $q_1, q_2$ as previously defined, since we have already shown as the kernels involved in the direct sum \eqref{sa_ext_q} map into each other unitarily.
    The domain of $\overline{q}_1$ and $\overline{q}_2$ can be written as (for details see \ref{App:dom}):
    \begin{align}
        \Dom{\overline{q}_1}&:= \{\psi \in \mathcal{H}_1\  | \ f(u) \partial_{u}\psi(u) \in \mathcal{H}_1,\, \lim_{u \to \pm \infty} \psi(u) =0\}  \\
         \Dom{\overline{q}_2}&:= \{\phi \in \mathcal{H}_2\  | \ 2f(v) \partial_{v}\phi(v)+\partial_{v} f(v) \phi(v) \in \mathcal{H}_2,\, \lim_{v \to \pm \infty} \sqrt{f(v)}\phi(v) =0\}
    \end{align}
   To prove that the domains above are correctly mapped into each other we need to show that if $\psi \in \Dom{\overline{q}_1} \Rightarrow  U \psi = f^{-1/2} \psi \in \Dom{ \overline{q}_2} $ and if $\phi \in \Dom{\overline{q}_2} \Rightarrow U^{-1} \phi = f^{1/2} \phi \in \Dom{\overline{q}_1}$. If $\psi \in \Dom{\overline{q}_1}$:
   \begin{align}
    \int \frac{du}{f(u)}f(u)^2 |\partial_u\psi(u)|^2 < + \infty \quad \text{and}\quad \lim_{u\to\pm\infty} \psi(u)=0
   \label{Xeqn18-2.20}
\end{align}
 Then, the condition $U \psi = f^{-1/2} \psi \in \Dom{ \overline{q}_2} $ is fulfilled since:
 \begin{align}
    \int_{\mathbb{R}} dv |2 f (f^{-1/2} \psi )' +f' f^{-1/2}\psi|^2=\int_{\mathbb{R}}dv |2 f^{1/2} \psi'|^2= 4 \int_{\mathbb{R}} dv \, f |\psi'|^2 < +\infty;
 \label{Xeqn19-2.21}
\end{align}
 and, immediately, follows that
 \begin{align}
     \lim_{v \to \pm \infty} \sqrt{f(v)}(U \psi)(v)=\lim_{v \to \pm \infty} \psi(v)=0.
 \label{Xeqn20-2.22}
\end{align}
On the other hand, if $\phi \in \Dom{\overline{q}_2}$:
 \begin{align}
     \int_{\mathbb{R}}dv |2 f(v) \partial_{v}\phi(v)+\partial_{v}f(v)\phi(v)|^2 < +\infty \quad \text{and}\quad  \lim_{v\to\pm\infty}\sqrt{f(v)}\phi(v)=0.
 \label{Xeqn21-2.23}
\end{align}
 Then, the condition $U \phi = f^{1/2} \phi \in \Dom{ \overline{q}_1} $ is satisfied since:
\begin{align}
    \int_{\mathbb{R}} \frac{du}{f}f^2 |(f^{1/2}\phi)'|^2 = \int_{\mathbb{R}}du  f |\tfrac{1}{2} f^{-1/2}f' \phi + f^{1/2} \phi '|^2= \int_{\mathbb{R}} du |\frac{1}{2} f' \phi + f \phi'|^2 < + \infty
\label{Xeqn22-2.24}
\end{align}
and, naturally, we conclude that
\begin{align}
    \lim_{u \to \pm \infty} (U^{-1}\phi)(u)=\lim_{u \to \pm \infty} \sqrt{f(u)}\phi(u)=0
\label{Xeqn23-2.25}
\end{align}

This clearly shows that the domains $\Dom{\overline{q}_1}$ and $\Dom{\overline{q}_2}$ map into each other, which in turn implies that the general domains $\Dom{q_{1}^{\lambda}},\Dom{q_{2}^{\lambda}}$ of the self-adjoint operators $q_1^{\lambda}, q_2^{\lambda}$ are mapped into each other by the unitary transformation $U$.
\end{proof}

\section{Dynamical equivalence}\label{Xsec4-3}
Another important consequence of unitarily equivalent Hilbert spaces is that the dynamics is actually the same. Given a Hamiltonian function $H(q,p)=T(p)+V(q)$, we can induce a Hamiltonian operator on two Hilbert spaces, $\mathcal{T}^{\lambda}$ and $\mathcal{K}^{\lambda}$, associated with the same $1$-dimensional GUP theory, namely $H_1=H(q_1,p_1)\in \mathcal{L}(\mathcal{H}_1)$ and $H_2=H(q_2,p_2)\in \mathcal{L}(\mathcal{H}_2)$. Suppose the Hamiltonian function is regular enough, then $UH_1U^{-1}=H_2$ and the two Hamiltonian operators have the same spectrum. \\
In the next subsection, we will discuss two relevant physical cases, showing explicitly the equivalence of the dynamics of the systems formulated within the triple $\mathcal{T}^{\lambda}$ and the triple $\mathcal{K}^{\lambda}$.
Examples of dynamical equivalence of certain physical phenomena formulated with respect to some specific deformed algebra (not necessarily GUP ones, strictly speaking) can be found scattered in the literature. For example in \cite{franchino2020casimir}, Casimir force analysis confirmed the dynamic equivalence of two different Snyder algebra realizations.

\subsection{An appropriate example: the harmonic oscillator}\label{Xsec5-3.1}

The harmonic oscillator represents one of the simplest yet fundamental physical systems that is worth discussing.
Unfortunately, a complete general analysis for a generic GUP algebra within the selected class is out of reach, since it is impossible to determine a general solution, for a generic $f(p)$, of the second-order differential equation corresponding to the eigenvalue problem of the system.
Therefore, we will limit our discussion by selecting a specific GUP algebra, namely:
\begin{align} \label{kmm_gup_algebra}
    [q,p]= i \hbar (1+ \beta p^2)
\end{align}
where $\beta$ is a positive parameter controlling the deformation---thus its physical scale---of the ordinary Heisenberg algebra.
The GUP scheme described by the algebra \eqref{kmm_gup_algebra} is the first model ever proposed, widely discussed in the literature, giving rise to a deformed uncertainty relation identical to that obtained in some string models \cite{kempf1994uncertainty,Konishi:1989wk}.

In both our GUP functional settings, the Hamiltonian of the one-dimensional harmonic oscillator is:
\begin{align}
    H= \frac{p^2}{2m}+\frac{1}{2}m \omega^2 q^2,
\label{Xeqn25-3.2}
\end{align}
with the usual meaning of the employed symbols.

For the triple $\mathcal{T}^\lambda$, the eigenvalue problem reads as:
\begin{align}
    \frac{u^2}{2m} \psi(u)- \hbar^2 m \omega^2 \beta u  (1+ \beta u^2)\frac{d\psi(u)}{du}- \frac{\hbar^2 m \omega^2}{2}(1+ \beta u^2) \frac{d^2 \psi(u)}{du^2}= E \psi(u)
\label{Xeqn26-3.3}
\end{align}

The complete exact solution of the problem was first discussed extensively in \cite{kempf1994uncertainty}, from which we can read the deformed eigenfunctions and the relative deformed energy spectrum:
\begin{align}
    &\psi(u)= \mathcal{N}_1 \frac{1}{(1+ \beta u^2)^{\left(\sqrt{s+r}\right)_n}} \;_2F_1\left(a_n, -n; c_n; \frac{1}{2}+ \frac{i \sqrt{\beta}u}{2}\right), \quad n \in \mathbb{{N}}, \label{kmm_eigenf}\\
    &E_n= \hbar \omega \left(n +\frac{1}{2}\right)\left(\frac{1}{4\sqrt{r}}+\sqrt{1+\frac{1}{16 r}}\right)+ \hbar \omega \frac{1}{4 \sqrt{r}}n^2, \label{kmm_en_spectrum}
\end{align}
where $\;_2F_1$ stands for the hypergeometric function, $\mathcal{N}_1$ is the proper normalization constant and we adopted the same notation used in \cite{kempf1994uncertainty}, namely:
\begin{align}
    \begin{aligned}
       &s_n=\frac{ E_n}{2 m \hbar^2 \omega^2 \beta}, \qquad r=\frac{1}{4 \beta^2 m^2 \hbar^2 \omega^2}, \qquad \left(\sqrt{s+r}\right)_n=\frac{1}{2}\left(n+\frac{1}{2}\right) +\frac{1}{4}\frac{\sqrt{4 +\beta^2 m^2 \hbar^2 \omega^2}}{\beta m \hbar \omega }, \\
       &a_n=-n -\frac{\sqrt{4 +\beta^2 m^2 \hbar^2 \omega^2}}{\beta m \hbar \omega }, \qquad c_n=1-2\left(\sqrt{s+r}\right)_n.
    \end{aligned}
\label{Xeqn27-3.6}
\end{align}

We will now proceed to exactly solve the harmonic oscillator problem for the triple $\mathcal{K}^{\lambda}$. Here, the eigenvalue problem reads as:
\begin{align}
    - \frac{\hbar^2 m \omega^2 }{2}(1+\beta v^2)^2 \frac{d^2 \varphi}{dv^2}- 2 \beta \hbar^2m \omega^2 v \frac{d\varphi}{dv}+\left(\frac{v^2}{2m}-\frac{\hbar^2 m \omega^2 \beta }{2}(1+ \beta v^2)-\frac{\hbar^2 m \omega^2 \beta^2 }{2}v- E\right) \varphi=0.
\label{Xeqn28-3.7}
\end{align}
The general solution for this second-order ODE can be written as:
\begin{align} \label{general_sol_Legendre}
    \varphi(v)=\frac{1}{\sqrt{1+ \beta v^2}}\left\{k_1 \mathcal{P}^{\mu}_{\gamma}(i \sqrt{\beta} v)+k_2 \mathcal{Q}_{\gamma}^{\mu}(i \sqrt{\beta} v)\right\},
\end{align}
where $\mathcal{P}^{\mu}_{\gamma}$ and $\mathcal{Q}^{\mu}_{\gamma}$ are respectively the associated Legendre functions of the first and second kind \cite{abramowitz1948handbook, olver2010nist} and the order $\gamma$ along with the degree $\mu$ are:
\begin{align}
    \gamma=\frac{1}{2}\left(-1+ \frac{\sqrt{4+ \beta^2 m^2 \hbar^2 \omega^2 }}{\hbar m \omega \beta }\right), \qquad \mu=\frac{\sqrt{1+2\beta E m}}{\hbar \omega m \beta }.
\label{Xeqn30-3.9}
\end{align}

Legendre functions can be conveniently expressed in terms of the hypergeometric functions \cite{abramowitz1948handbook, olver2010nist}:
\begin{align} \label{index_Legendre_f}
   \mathcal{P}^{\mu}_{\gamma}(i \sqrt{\beta}v)=&
  \left( \frac{1 + i \sqrt{\beta} \, p}{1 - i \sqrt{\beta} \, v} \right)^{\mu / 2}
  \frac{{}_2F_1\left( -\gamma, \gamma + 1; 1 - \mu; \frac{1}{2}- \frac{i \sqrt{\beta} \, v}{2} \right)}{\Gamma(1 - \mu)},  \\
  \mathcal{Q}^{\mu}_{\gamma}(i \sqrt{\beta} v)=&\frac{\pi}{2} \csc(\mu \pi) \biggl[
  \left(
    \frac{1 + i \sqrt{\beta} \, v}{1 - i \sqrt{\beta} \, v}
  \right)^{\mu/2}
  \cos(\mu \pi)
  \;
  \frac{
    {}_2F_1\left( -\gamma, \gamma + 1; 1 - \mu; \frac{1}{2} - \frac{i \sqrt{\beta} \, v}{2} \right)}{  \Gamma(1 - \mu)}   \nonumber \\
 & -
  \left(
    \frac{(1 - i \sqrt{\beta} \, p)^{\mu/2}}{(1 + i \sqrt{\beta} \, v)^{\mu/2}}
  \right)
  \frac{\Gamma(1 + \mu + \gamma)}{\Gamma(1 - \mu + \gamma)}
  \frac{\;_2F_1\left( -\gamma, \gamma + 1; 1 + \mu; \frac{1}{2} - \frac{i \sqrt{\beta} \, v}{2}  \right)}{\Gamma(1 + \mu)}\biggr]
 \end{align}
These identities are valid in the complex unit disk, but they can be analytically continued to the cut complex plane, suitably avoiding the singular points $(-1,1, \infty)$ of the Legendre functions.
In particular, being the complex variable $z=i \sqrt{\beta} v$, as $v$ varies, we move along the imaginary axis, safely avoiding all the singular points for all the real values of $v$.

Associated Legendre functions are not square integrable functions in general; therefore, we need to impose some conditions to deal with proper physical states.
To do so, we asymptotically expand the function at $v=\pm \infty$.
Since the asymptotic behavior at both endpoints is the same, up to overall signs of the coefficients, we can restrict our discussion to the expansion at $v=\infty$:
\begin{align}
    \varphi(v) \underset{v \to + \infty} {\sim }\left(C_{1}^{\gamma,\mu}+C_{3}^{\gamma,\mu}+C_{5}^{\gamma,\mu}\right)v^{-2 -\gamma}+\left(C_{2}^{\gamma,\mu}+C_{4}^{\gamma,\mu}+C_{6}^{\gamma,\mu}\right)v^{-1 +\gamma},
\label{Xeqn31-3.12}
\end{align}
where $C_i^{\gamma,\mu}$ are the coefficients of the expansion, depending on the indices $\gamma, \mu$.
It is easy to verify that the $L^2$-integrability condition $2+\gamma > 1/2$ is always satisfied. However, the condition $1-\gamma > 1/2$ is always violated, except for a specific relation among the physical constants involved. Since imposing such a constraint would not have any physical meaning, we must require that every coefficient of the $p^{-1+\gamma}$ term vanishes individually to ensure that the state is physically acceptable.
These coefficients read as:
\begin{align*}
&C_{2}^{\gamma,\mu}=\frac{
  2^{\gamma} \, c_1 \, e^{i \pi \mu / 2} \, \Gamma\left( \tfrac{1}{2} + \gamma \right)
}{\sqrt{\pi} \, \sqrt{\beta} \, \Gamma(1 - \mu + \gamma)}, \\
&C_4^{\gamma,\mu}=\frac{
  2^{-1 + \gamma} \, c_2 \, \sqrt{\pi} \, \cot(\pi \mu) \, \Gamma\left( \tfrac{1}{2} + \gamma \right)}{\sqrt{\beta} \, \Gamma(1 - \mu + \gamma) }, \\
&C_6^{\gamma,\mu}=- \frac{
  2^{-1 + \gamma} \, c_2 \, \sqrt{\pi} \, \csc(\pi \mu) \, \Gamma\left( \tfrac{1}{2} + \gamma \right)}{ \sqrt{\beta} \, \Gamma(1 - \mu + \gamma)}.
\end{align*}
The only condition that ensures the vanishing of all three terms is the divergence of the common denominator, i.e., we need to ask that the argument of the Gamma functions in the denominator is a negative integer number, namely $1+\gamma-\mu=-n$.
By using expressions \eqref{index_Legendre_f}, we can solve for $E$ and obtain the energy spectrum of the GUP harmonic oscillator compatible with the square integrability requirement:
\begin{align}
    E_n=\hbar \omega \left(n+\frac{1}{2}\right)\left(\frac{1}{2}\hbar m \omega \beta + \frac{1}{2}\sqrt{4 + m^2 \beta^2 \hbar^2 \omega^2 }\right)+\frac{1}{2} \hbar^2\omega^2 \beta m n^2.
\label{Xeqn32-3.13}
\end{align}
This is exactly the energy spectrum in \eqref{kmm_en_spectrum}, as can be verified using the explicit expression of the $r$ quantity.

With this quantization condition, the general solution \eqref{general_sol_Legendre} can be written as:
\begin{align}
    \varphi(v)=\frac{1}{\sqrt{1+ \beta v^2}}\left(\frac{c_1+c_2\frac{\pi}{2}\cot{(\mu_n \pi)}}{\Gamma(1- \mu_n)}\right) \left( \frac{1 + i \sqrt{\beta} \, v}{1 - i \sqrt{\beta} \, v} \right)^{\mu_n / 2}
  {}_2F_1\left( -\gamma, \gamma + 1; 1 - \mu_n; \frac{1}{2}- \frac{i \sqrt{\beta} \, v}{2} \right),
\label{Xeqn33-3.14}
\end{align}
where $\mu_n$ denotes the $\mu$ quantity taking into account the energy quantization.\\
It is not difficult to verify that the obtained wave function correctly lies in the domain of the self-adjoint operator $H$---as it should---, which in turn can be easily obtained starting from the domains of the $q^{\lambda}_i,p_i$ operators in both formulations.\\
We can now use one of the connection formulas concerning the hypergeometric function, precisely \cite{olver2010nist}:
\begin{align}
    \begin{split}
        {}_2F_1(a,b;c;z)
= &\frac{\Gamma(c)\,\Gamma(c - a - b)}{\Gamma(c - a)\,\Gamma(c - b)}\;
  {}_2F_1\bigl(a,b;\,a + b + 1 - c;\,1 - z\bigr)
 \\
&+\frac{\Gamma(c)\,\Gamma(a + b - c)}{\Gamma(a)\,\Gamma(b)}\;
  (1 - z)^{c - a - b}\;
  {}_2F_1\bigl(c - a,\;c - b;\;1 + c - a - b;\;1 - z\bigr).
    \end{split}
\label{Xeqn34-3.15}
\end{align}
If applied in our context, by keeping in mind the relation $1-\mu+\gamma=-n$ and the symmetry between the arguments $a,b$ of the hypergeometric function, we can write:
\begin{align}
\varphi(v)=\frac{2^{\mu_n}\Gamma(\mu_n)}{{\sqrt{1+ \beta v^2}}}\left(\frac{c_1+c_2\frac{\pi}{2}\cot{(\mu_n \pi)}}{\Gamma(1- \mu_n)}\right)\frac{1}{\left(1+ \beta v^2\right)^{\mu_n/2}} {}_2F_1\left(-\mu_n-\gamma, -n; 1 - \mu_n; \frac{1}{2}+\frac{i \sqrt{\beta} v}{2}\right)
\label{Xeqn35-3.16}
\end{align}
By using the explicit form of $\gamma$ and $\mu_n$, it is straightforward to show that the following identities hold:
\begin{align}
    \mu_n-\gamma= a_n, \quad 1- \mu_n=c_n, \quad \frac{\mu_n}{2}=\left(\sqrt{s+r}\right)_n.
\label{Xeqn36-3.17}
\end{align}
This allows us to finally rewrite the general solution \eqref{general_sol_Legendre} as:
\begin{align}
\varphi(v)=\frac{\mathcal{N}^n_2}{{\sqrt{1+ \beta v^2}}}\frac{1}{\left(1+ \beta v^2\right)^{\left(\sqrt{s+r}\right)_n}} {}_2F_1\left(a_n, -n; c_n; \frac{1}{2}+\frac{i \sqrt{\beta} v}{2}\right),
\label{Xeqn37-3.18}
\end{align}
from where it can be immediately recognized as the mapping via the unitary operator $U$ of the eigenfunction $\psi(u)$ in \eqref{kmm_eigenf}, as it should be.

\subsection{An inappropriate example: the free fall in the Newtonian regime}\label{Xsec6-3.2}
We now wish to analyze another simple case to demonstrate the correct behavior of the transformation of the Schr{\" o}dinger equation under equivalent Hilbert spaces: the free-fall scenario. Although this example is not ideal, as the Hamiltonian has a continuous spectrum, we can nonetheless carry out the analysis for a generic function $f$, ultimately showing that the solutions indeed transform under the unitary map \eqref{u-map}.

\medskip

The free fall is described by $H=\frac{p^2}{2m}+mgq$.
For the triple $\mathcal{T}^{\lambda}$, we obtain:
\begin{align}
    \frac{u^2}{2m}\psi(u)+i\hbar mgf(u)\frac{d\psi}{du}=E\psi(u),
\label{Xeqn38-3.19}
\end{align}
which can be solved by separation of variables, obtaining
\begin{align}
    \psi_E(u)=\exp\left(-\frac{i}{2\hbar m^2g}\int\frac{2mE-u^2}{f(u)}dx \right).
\label{Xeqn39-3.20}
\end{align}
In space $\mathcal{K}^{\lambda}$, the equation becomes
\begin{align}
    \frac{v^2}{2m}\varphi(v)+i\hbar mgf(v)\frac{d\varphi}{dv}+\frac{1}{2}i\hbar mg \varphi(v)\frac{df}{dv}=E\varphi(v).
\label{Xeqn40-3.21}
\end{align}
Using again the separation of variables, we obtain
\begin{align}
    \frac{1}{\varphi(v)}\frac{d\varphi}{dv}=-\frac{1}{2f(v)}\frac{df}{dv}+\frac{1}{2i\hbar  m^2 g}\frac{2mE-v^2}{f(v)},
\label{Xeqn41-3.22}
\end{align}
with solution
\begin{align}
    \varphi_E(v)=\frac{1}{\sqrt{f(v)}}\exp\left(-\frac{i}{2\hbar m^2g}\int\frac{2mE-v^2}{f(v)}dv \right).
\label{Xeqn42-3.23}
\end{align}
A proper way to analyze this transformation is in terms of linear functionals. Whenever a linear transformation $L:\mathcal{H}_1\to\mathcal{H}_2$ exists, we can map linear functionals by the pullback. Namely, let $F:\mathcal{H}_2\to\C$ be a linear functional, then $L^*F:\mathcal{H}_1\to\C$, where $(L^*F)[\psi]=F[L\psi]$ defines a linear functional on $\mathcal{H}_1$.\\
For a finer analysis, we must consider the Gel'fand triple $\mathcal{D}\subset L^2\simeq (L^2)^*\subset\mathcal{D}^*$, where $\mathcal{D}$ is the set of smooth maps with compact support $C_c^{\infty}(\R)$ and $\mathcal{D}^*$ its continuous dual space \cite{gel2014generalized}. Indeed, the solutions to both the Schr\"odinger equations are continuous functions with finite sup-norms, then they act on $\mathcal{D}$ as linear functionals. On $\mathcal{H}_1$, we can define a family of functionals $F_E:\mathcal{D}\subset L^2\left(\R,\frac{du}{f(u)}\right)\to\C$ by
\[F_E[\psi]=\int_{\R}\psi_E(u)\psi(u)\frac{du}{f(u)},\] while, on the space $\mathcal{H}_2$, we have $G_E:\mathcal{D}\subset L^2(\R,dv)\to\C$ defined by
\[G_E[\phi]=\int_{\R}\varphi_E(v)\phi(v)dv.\]

In the analyzed case, $F_E$ is the pullback of $G_E$ via the unitary map $U$ in \eqref{u-map}, namely $U^*G_E=F_E$. Indeed
\begin{align}
    \begin{split}
        G_E[U\psi]&=\int_{\R}\varphi_E(v)\frac{1}{\sqrt{f(v)}}\psi(v)dv=\int_{\R}\frac{1}{\sqrt{f(v)}}\psi_E(v)\frac{1}{\sqrt{f(v)}}\psi(v)dv\\
    &=\int_{\R}\psi_E(u)\psi(u)\frac{du}{f(u)}=F_E[\psi].
    \end{split}
\label{Xeqn43-3.24}
\end{align}

\section{Comment on polymer-like approach and inequivalent representations}\label{Xsec7-4}
The construction we developed allows us to shed light on the relationship between the usual functional GUP setting represented by the triple $\mathcal{T}^{\lambda}$ and another functional configuration sometimes employed in treating GUP theories, namely a \emph{polymer quantum mechanics}-like formulation.
Polymer quantum mechanics (PQM) is a particular formulation of quantum mechanics obtained through a non-unitarily equivalent representation of the Weyl-Heisenberg algebra, constructed to mimic, on a different level, the loop quantization procedure developed for cosmology \cite{corichi2007hamiltonian, corichi2007polymer}. Remarkably, the formalism succeeds in reproducing a granular structure of space, introducing a fundamental scale which can be associated with the Planck scale.
The functional structure of the theory with respect to the construction of the Hilbert space and all that follows is very different from the GUP framework.
In particular, in the PQM context, according to the chosen representation, it is not possible to define simultaneously both the momentum and position operators as self-adjoint operators. \\
One of the most common choices is to work in a representation in which we can define properly the position operator, but it is not possible to define the momentum operator.
In a similar setting, the position operator can be represented as a differential one, while the momentum operator has to be regularized and represented as an approximation relying on the graph structures, which can be introduced properly in the theory. In general, this will lead to a proper momentum operator represented by means of a trigonometric function.
Inspired by this construction, analogous representation has been used for GUP theories as well, adapted to the functional structure inherent to this class of theories (see e.g. \cite{Barca:2021epy}).
We can use the scheme developed for the definition of the second triple above to properly recast this GUP formulation.
Let us consider a new triple $(\mathcal{H}_{poly}, q_{poly},p_{poly})$, where $\mathcal{H}_{poly}:= L^2(\mathbb{R},dz)$ and $q_{poly}:= q_{poly}(Q,P)$, $p_{poly}:= p_{poly}(Q,P)$, with $Q,P$ the self-adjoint auxiliary operators already defined.
To mimic the PQM setting, we fix the following representation of the physical operators $q_{poly}, p_{poly}$:
\begin{align}
    (q_{poly}\psi)(z)=i \hbar \partial_{z} \psi(z),\ \ \
    (p_{poly}\psi)(z)=g(z)\psi(z).
\label{Xeqn44-4.1}
\end{align}
Clearly, $q_{poly}$ is acting as a differential operator, while $p_{poly}$ is acting as a multiplicative operator, with a precise functional form. By requiring this representation to satisfy the GUP algebra, we can easily determine the general form of the $g(z)$ function:
\begin{align}
    \int_{0}^{g(z)}\frac{ds}{f(s)}=z \, \Rightarrow g(z)=h^{-1}\left(z\right),
\label{Xeqn45-4.2}
\end{align}
where we have set:
\begin{align}
    h(s):= \int_0^s \frac{dt}{f(t)}
\label{Xeqn46-4.3}
\end{align}
and we are denoting with $h^{-1}$ the inverse function of $h$.
With respect to the auxiliary operators $Q,P$ it is clear that we can express our physical operators $q_{poly}, p_{poly}$ as:
\begin{align}
    q_{poly}(Q,P)=Q, \quad p_{poly}(Q,P)=g(P).
\label{Xeqn47-4.4}
\end{align}
To verify that this formulation is unitarily equivalent to the one represented by the triple $\mathcal{T}^{\lambda}$, we should identify a new proper unitary transformation $U_{poly}$ which, along with its inverse, properly maps the action and the domains of the operators between the two Hilbert spaces.

Nevertheless, according to what we have demonstrated in Proposition {\ref{core_props}},  since we already know that such transformation has to guarantee the same functional properties of the operator $q_1^{\lambda}$ and the operator $q_{poly}$, we can directly conclude that no GUP theories with a minimal uncertainty in position different from zero can be unitarily equivalent to the polymer-like formulation.
Indeed, the operator $q_{poly}(Q,P)=Q$, in the Hilbert space $\mathcal{H}_{poly}$, is an essentially self-adjoint operator, with a unique self-adjoint extension represented by the operator $q_{poly}^{\dagger}$ defined on the following domain:
\begin{align}
    \Dom{q_{poly}^{\dagger}}:= \{\psi \in \mathcal{H}_{poly} \, | \, \partial_z \psi(z) \in \mathcal{H}_{poly} \}.
\label{Xeqn48-4.5}
\end{align}
This necessarily means that ${\rm Ker}(q_{poly}^{\dagger}\pm i \mathbb{I}_{\mathcal{H}_{poly}})=\{0\}$. On the other hand, if we are dealing with GUP theories represented on $L^2\left(\R,\frac{du}{f(u)}\right)$ which allow a non-zero minimal uncertainty in position, the position operator $q_1$ will not be essentially self-adjoint but rather it will admit infinitely many self-adjoint extensions. This correspondence between the physical property of the presence of a minimal uncertainty in a given observable and the mathematical feature of the loss of essential self-adjointness of the associated operator is a well-established fact in the literature \cite{Kempf:1994su, Martin:2009qzj}. Technically this means that $\dim {\rm Ker}(q_1^{\dagger}\pm i \mathbb{I}_{\mathcal{H}_1})\neq 0$ or, in our case, that $\dim {\rm Ker}(q_1^{\dagger}\pm i \mathbb{I}_{\mathcal{H}_1})=1$, being the $q_1$ operator a first-order differential operator.
It is thus clear that no unitary transformation can exist in this context to map correctly the first triple in the second triple, implying that these two formulations cannot be considered as physically equivalent. \\
From a physical point of view, this is not surprising. Indeed, the two different functional properties of the position operators involved imply contrasting localization properties, which clearly cannot be reconciled.\\
From what we have said, it is clear that GUP theories formulated in $L^2\left(\R,\frac{du}{f(u)}\right)$, for which the position operator turns out to be essentially self-adjoint \emph{could} be unitarily equivalent to the polymer-like formulation, since in principle, in this case, a unitary transformation could exist.
Despite that, since the most physically relevant GUP theories are the ones that lead to the loss of point-wise localization, i.e. theories with a not essentially self-adjoint position operator, we will not delve further into this scenario.
However, what we have just discussed possibly provides a violation of the Stone-von Neumann theorem for this kind of theories. Indeed, even if $\mathcal{T}^{\lambda}$ and $(\mathcal{H}_{poly}, q_{poly},p_{poly})$ are Hilbert spaces associated with the same one-dimensional GUP theory, they may not be unitarily equivalent, as discussed above, thus, providing that a generalization of the Stone-von Neumann theorem for Generalized Uncertainty Principle theories could not exist in accordance with our definitions of unitary equivalence.

\section{Conclusion}\label{Xsec8-5}

In this work, we have examined the delicate and fundamental issue of unitary equivalence within the context of GUP theories, limiting ourselves to the more manageable one-dimensional case.
Unitary equivalence plays a crucial role in quantum physics, as it provides the framework to determine whether or not two distinct formulations actually describe the same physical content.
In ordinary quantum mechanics, the well-known Stone-von Neumann theorem clearly specifies all the conditions under which different realizations of the Heisenberg algebra are unitarily equivalent.
Given this, it is quite natural to raise the same question within the framework of GUP models.
Indeed, given a deformed Heisenberg algebra, representing it properly requires selecting a suitable Hilbert space in which the conjugate operators $q, p$ act, defined on a dense domain where they are self-adjoint.
This procedure, which allows us to associate a GUP theory with a triple $(\mathcal{H}, q, p)$, is not unique, and this raises the issue of unitary equivalence.\\
To address this matter, as a first step, we have provided a formal definition for a GUP theory as in Def.~\ref{def:gup-th} and stated two requirements that clarify what we mean by equivalence. Namely, given two GUP formulations, we asked for the correspondence of the functions $f_i$ associated with the algebras with respect to their spectrum $\sigma(f_i)$ and the existence of a unitary mapping between the two Hilbert spaces involved, such that the different realizations of the conjugate operators are related to each other.\\
Within this general framework, we focused our investigation on the unitary equivalence between two of the main GUP constructions found in the literature.
The first is that introduced by Kempf, Mangano and Mann, where the Hilbert space is equipped with a deformed Lebesgue measure to ensure the symmetry of the position operator. This is perhaps the most ``natural'' realization for a GUP theory. The other commonly used approach is based on the standard Hilbert space, together with the introduction of operators $Q, P$ which satisfy the conventional Heisenberg algebra. These operators should be regarded as auxiliary variables, in terms of which the physical ones $q, p$ can be expressed.\\
We were able to discuss in detail all the relevant functional properties of these frameworks, demonstrating that they both fit the definition of GUP theory we provided, and we explicitly constructed a unitary transformation that correctly relates the two constructions, ensuring that the self-adjoint extensions (whether unique or infinitely many) of the conjugate operators, along with their domains, are properly mapped by the transformation we established.
This allows us to conclude that these two principal formulations, under the imposed conditions--particularly the self-adjointness not only of the physical operators $q, p$ but also of the auxiliary operators $Q, P$---are equivalent and describe the same quantum phenomena.
In other words, this provides a concrete example of a quantum GUP-canonical transformation represented as a unitary map.\\
As a final check, we tested our conclusions by analyzing the dynamics of two very simple yet significant physical systems: the quantum harmonic oscillator and the free fall of a quantum particle. We solved for the spectrum of the Hamiltonian in both frameworks, along with its eigenfunctions, and directly demonstrated how they are related through the unitary transformation we constructed. While the first case is solved for a specific choice of the function $f$, the second allows for a general computation, although it requires the use of the language of Gel'fand triples to rigorously prove the unitary equivalence of the system's dynamics.\\
Finally, we discussed a counterexample, namely a situation in which two alternative constructions of the same GUP algebra are not equivalent.
Specifically, we showed how the ``natural'' GUP realization, commonly adopted in the study of GUP dynamics, is not unitarily equivalent---at least for those classes of GUP theories that lead to the emergence of a minimal length---to the polymer-like GUP construction.
This approach is inspired by Polymer Quantum Mechanics (PQM) and, although in quite a different context, essentially requires the position operator to act as an ordinary derivative and the momentum operator to have a modified multiplicative action in momentum representation. By demonstrating, for the class of GUP theories mentioned above, that the functional properties of the position operators $q$ and $q_{\text{poly}}$ are necessarily distinct, we were able to conclude that no unitary transformation can exist to connect the two frameworks. Indeed, if such a mapping existed, it would have to preserve the functional characteristics of the position operators in both cases. \\
The importance of this example, ultimately, lies in the fact that it supports the idea that a generalization of the Stone-von Neumann theorem for GUP theories may not hold, at least with respect to the definition of equivalence we have adopted, since the same GUP algebra can admit representations that are not unitarily equivalent. \\
In summary, this work clarifies, in the one-dimensional context, the construction of unitarily equivalent GUP theories, providing a set of criteria that must be satisfied in order to ensure that we are dealing with the same physical content. In turn, if this is the case, it proves that quantum GUP-canonical transformations exist and opens the way for their more detailed investigation.

\section*{Acknowledgement}
M.B. is supported by the MUR FIS2 Advanced Grant ET-NOW (CUP:~B53C25001080001) and by the INFN TEONGRAV initiative.

\appendix
\section{Construction of the domains}
\label{App:dom}

In the proof of Proposition {\ref{core_props}}, to show how the unitary map $U$ properly transforms the self-adjointness domains of the position operators $q_1^{\lambda},q_2^{\lambda}$, we used the explicit form of the domains of the closed operators $\overline{q}_1, \overline{q}_2$. Here we show how to construct them.\\
Let us consider the operators $q_1$ and $q_2$ defined on some proper dense domain of their respective Hilbert spaces, on which they turn out to be symmetric, e.g. the space $C^{\infty}_c(\mathbb{R})$ of all smooth, compactly supported functions. For symmetric operators, to easily determine their closure, we can compute their bi-adjoint,  since $\overline{q}=q^{\dagger\dagger}$ \cite{moretti2013spectral}. \\
First, we can construct the adjoint of these operators directly from the definition. For the operator $q_1$, we have, by definition of adjoint operator:
\begin{align}
    \begin{split}
        &\psi \in \Dom{q_1^\dagger} \subset L^2\left(\mathbb{R}, \frac{du}{f(u)}\right)  \iff \exists  \chi \in L^2\left(\mathbb{R}, \frac{du}{f(u)}\right): \\
  & i\hbar \int_{\mathbb{R}} du \, \psi^{*}(u) \frac{d\varphi}{du}=\int_{\mathbb{R}} \frac{du}{f(u)}\chi^{*}(u) \varphi(u), \, \forall \varphi \in \Dom{q_1},
    \end{split}
\label{Xeqn49-A.1}
\end{align}
where, by definition, $q_1^{\dagger}\psi=\chi$.
Since $\Dom{q_1}=C^{\infty}_c(\R)\subset L^2\left(\mathbb{R}, \frac{du}{f(u)}\right)$, the condition above, as formulated, is related to the weak derivative. In particular we can write:
\begin{align}
   \psi  \in \Dom{q_1^\dagger} \iff  i \hbar f(u) \partial_{u} \psi (u) = \chi(u)\in L^2\left(\mathbb{R}, \frac{du}{f(u)}\right).
\label{Xeqn50-A.2}
\end{align}
where $\partial_u\psi$ is the weak derivative of $\psi$ as defined below.
\begin{definition}    Let $\psi$ $\in$ $L^2\left(\mathbb R,\frac{du}{f(u)}\right)$, we say that $\xi\in L^2\left(\mathbb R,\frac{du}{f(u)}\right)$ is its \emph{weak derivative} if
        \begin{align*}
        \int_\mathbb{R}\psi(u)\frac{d\varphi}{du}du=-\int_\mathbb{R}\xi(u)\varphi(u)du,\quad \forall\varphi\in C^{\infty}_c(\mathbb{R}).
    \end{align*}
    The weak derivative, if it exists, is unique and we indicate it using the {standard symbol for differentiation} $\xi:=\partial_u\psi=\frac{d\psi}{du}=\psi'$.
\label{Xenun5-4}
\end{definition}
    This definition of weak derivative for functions in $L^2\left(\mathbb R,\frac{du}{f(u)}\right)$ is well-posed because both sides of the equation are finite, indeed:
    \begin{align}
        \begin{split}
            &\left|\int_\mathbb{R}{\psi^{*}(u)}\frac{d\varphi}{du}du\right |\leq \max_{u\in\mathrm{supp}(\varphi)}f(u) \left|\bra\psi,\varphi'\ket_{L^2\left(\mathbb{R},\frac{du}{f(u)}\right)}\right|<+\infty,\\
            &\left|\int_\mathbb{R}{\xi^{*}(u)}\varphi(u)du\right|\leq\max_{u\in\mathrm{supp}(\varphi)}f(u) \left|\bra\xi,\varphi\ket_{L^2\left(\mathbb{R},\frac{du}{f(u)}\right)}\right|<+\infty.
        \end{split}
    \label{Xeqn51-A.3}
\end{align}

Thus:
\begin{align} \label{adj_q1_dom}
    \Dom{q_1^{\dagger}}=\left\{\psi(u) \in L^2\left(\mathbb{R}, \frac{du}{f(u)}\right)\  | \  f(u) \partial_{u} \psi (u) \in  L^2\left(\mathbb{R}, \frac{du}{f(u)}\right) \right\}.
\end{align}
Following the same steps, for the adjoint $q_2^\dagger$ of the position operator $q_2$ in $\mathcal{H}_2$, we obtain the following domain:
\begin{align} \label{adj_q2_dom}
     \Dom{q_2^{\dagger}}=\left\{\psi(v) \in L^2\left(\mathbb{R}, dv\right)\ | \  2 f(v) \partial_{v} \psi (v)+\partial_v f \psi (v) \in  L^2\left(\mathbb{R}, dv\right) \right\}.
\end{align}
We can construct the adjoint of $q_1^{\dagger}, q_2^{\dagger}$, that is the bi-adjoint of $q_1,q_2$, reiterating the definition above, in view of the domains \eqref{adj_q1_dom} and \eqref{adj_q2_dom}. \\
In the Hilbert space $\mathcal{H}_1$ we have that:
\begin{align}
    \begin{split}
        &\phi \in \Dom{q_1^{\dagger\dagger}} \subset L^2\left(\mathbb{R}, \frac{du}{f(u)}\right)  \iff \exists  \chi \in L^2\left(\mathbb{R}, \frac{du}{f(u)}\right): \\
  & i \hbar \int_{\mathbb{R}} du \, \phi^{*}(u) \frac{d\psi}{du}=\int_{\mathbb{R}} \frac{du}{f(u)}\chi^{*}(u) \psi(u), \, \forall \psi \in \Dom{q_1^\dagger},
    \end{split}
\label{Xeqn54-A.6}
\end{align}
Given that now $\psi \in \Dom{q_1^\dagger}$, no specific boundary conditions hold; therefore, we cannot recognize directly the integral equality above as the definition for the weak derivative.
We can perform {an integration by parts}, noticing that $\psi \in \Dom{q_1^\dagger}$ and asking for $\phi$ to admit weak derivative in $L^2\left(\mathbb{R}, \frac{du}{f(u)}\right)$ (the precise meaning of this {integration by parts} is explained later):
\begin{align}
\label{int_parts_1}
  \phi(u)^{*}\psi(u)\bigg|^{+\infty}_{-\infty}- \int_{\mathbb{R}} du \, \psi(u) \partial_{u} \phi^{*}(u)=\frac{1}{i \hbar}\int_{\mathbb{R}} \frac{du}{f(u)}\chi^{*}(u) \psi(u)
\end{align}
By demanding the vanishing of the boundary term and the equality of the two integrals, we obtain the domain of the operator $q_1^{\dagger\dagger}$:
\begin{align} \label{biadj_q1_dom}
\begin{aligned}
    \Dom{q_1^{\dagger\dagger}}:= &\left\{\phi(u) \in L^2\left(\mathbb{R}, \frac{du}{f(u)}\right)\ | \  f(u) \partial_{u} \phi (u) \in  L^2\left(\mathbb{R}, \frac{du}{f(u)}\right), \lim_{u \to \pm \infty}  \phi(u)=0\right\}.
\end{aligned}
\end{align}
The treatment of the integral and the boundary term requires some subtleties. We need to characterize the properties of the functions in $\Dom{q_1^\dagger}$. We start noticing that $L^2\left(\mathbb{R}, \frac{du}{f(u)}\right)\subset L^2\left([a,b], \frac{du}{f(u)}\right)$ for any $a<b\in \R$, by monotonicity of the integral and positivity of $f$. Moreover, since $f$ is bounded by below by a positive number, i.e. $f\geq c>0$, it holds
\begin{align}
    \int_U\frac{du}{f(u)}\leq \frac{1}{c}\int_Udu, \text{ for every measurable subset $U\subset\R$.}
\label{Xeqn57-A.9}
\end{align}
This implies that the deformed measure is dominated by the Lebesgue one and the compact interval $[a,b]$ has finite volume. As a consequence $L^2\left([a,b], \frac{du}{f(u)}\right)\subset L^1\left([a,b], \frac{du}{f(u)}\right)$. Thus, if $\psi\in\Dom{q_1^\dagger}$, then $\psi'\in L^1([a,b],du)$ for any $a<b\in \R$, and so its antiderivative $F$ is absolutely continuous on $[a,b]$ by the fundamental theorem of Lebesgue calculus:
\begin{align}
    F(u)=\int_a^u\psi'(s)ds.
\label{Xeqn58-A.10}
\end{align}
Moreover $F$ is continuous on the whole $\R$ and $F(u)=\psi(u)-\psi(a)$ almost everywhere on $\R$. In particular $F$ is uniformly continuous, indeed:
\begin{align}
\begin{split}
    |F(u)|&=\left|\int_a^u\psi'(s)ds\right|=\left|\int_a^u(f(s)\psi'(s))1\frac{ds}{f(s)}\right|=|\bra1,f\psi'\ket_{L^2\left([a,u],\frac{ds}{f(s)}]\right)}|\\
    &\leq\left|||f\psi'||_{L^2\left([a,u],\frac{ds}{f(s)}]\right)}||1||_{L^2\left([a,u],\frac{ds}{f(s)}]\right)}\right|\leq||f\psi'||_{\mathcal{H}_1}\sqrt{\left |\int_a^u\frac{ds}{f(s)}\right|}\\
    &\leq \frac{||f\psi'||_{\mathcal{H}_1}}{\sqrt{c}}\sqrt{|u-a|}.
\end{split}
\label{Xeqn59-A.11}
\end{align}
We can then choose $\psi$ to be uniformly continuous and absolutely continuous on every compact interval by changing it on a set of measure zero. We can now prove that $\psi$ is a bounded function. Consider the condition $\psi\in \Dom{q_1^\dagger}$ written as:
\begin{align}
    \int_\R \left|\frac{\psi(u)}{\sqrt{f(u)}}\right|^2du<+\infty, \qquad \int_\R \left|\sqrt{f(u)}\psi'(u)\right|^2du<+\infty.
\label{Xeqn60-A.12}
\end{align}
We have then two functions $g_1(u)=\frac{\psi(u)^*}{\sqrt{f(u)}}$ and $g_2(u)=\sqrt{f(u)}\psi'(u)$ living in $L^2(\R,du)$, and so in $L^2([0,a],du)$, thus for these two functions the Cauchy-Schwartz inequality holds:
\begin{align}
    |\bra g_1, g_2\ket_{L^2([0,a],du)}|\leq ||g_1||_{L^2([0,a]),du)}||g_2||_{L^2([0,a],du)}\leq||g_1||_{L^2(\R,du)}||g_2||_{L^2(\R,du)}=L <+\infty.
\label{Xeqn61-A.13}
\end{align}
But, this product is
\begin{align}
    \bra g_1, g_2\ket_{L^2([0,a],du)}=\int_0^{a}\frac{\psi(u)}{\sqrt{f(u)}}\sqrt{f(u)}\psi'(u)du=\int_0^{a}\psi(u)\psi'(u)du=\frac{1}{2}(\psi(a)^2-\psi(0)^2).
\label{Xeqn62-A.14}
\end{align}
Hence, for any $a\in \R$, the function is bounded, namely $|\psi(a)|\leq \sqrt{2L+|\psi(0)|^2}$. We can now give a precise meaning to the integration by parts in Eq.{\eqref{int_parts_1}}: let $\psi,\phi\in\Dom{q_1^\dagger}$, then they are absolutely continuous functions with $L^1$ weak derivative on every compact interval, as previously proved. Let us suppose $\phi\in\Dom{q_1^{\dagger\dagger}}$, which is contained in $\Dom{q_1^\dagger}$, because, for any symmetric operator, holds
\begin{align}
    \Dom{q_1}\subseteq\Dom{q_1^{\dagger\dagger}}\subseteq\Dom{q_1^{\dagger}}.
\label{Xeqn63-A.15}
\end{align}
Considering the integral
\begin{align}
    \int_\R\phi^*(u)\frac{d\psi}{du}du=\lim_{l\to+\infty}\int_{-l}^l\phi^*(u)\frac{d\psi}{du}du,
\label{Xeqn64-A.16}
\end{align}
for each fixed $l$, the necessary and sufficient conditions to apply the integration by parts on a compact interval ($\phi$ is absolutely continuous and $\psi'$ is $L^1$ on the interval) are satisfied. Then we can integrate by parts:
\begin{align}
    \lim_{l\to+\infty}\int_{-l}^l\phi^*(u)\frac{d\psi}{du}du=\lim_{l\to+\infty}\left(\phi(l)\psi(l)-\phi(-l)\psi(-l)-\int_{-l}^l\psi(u)\frac{d\phi^*}{du}du\right)
\label{Xeqn65-A.17}
\end{align}

Now, the product of a bounded function times a function that vanishes at infinity still vanishes at infinity, hence the request {$\lim_{u\to \pm \infty}\phi(u)=0$} is enough to ensure the vanishing of the boundary term. More compactly, we can write:
\begin{align}
    \Dom{q_1^{\dagger\dagger}}=\Dom{q_1^{\dagger}}\cap C_0(\R),
\label{Xeqn66-A.18}
\end{align}
where $C_0(\R)$ is the space of continuous functions over $\R$ that vanish at infinity.

It is worth noting that, in case of a constant $f$, $\Dom{q_1^{\dagger}}$ is just the Sobolev space $H^1(\R)$, in which the functions automatically vanishes at infinity, ensuring self-adjointness of the operator $q_1^{\dagger}$. One could ask if the request of $f$ to be smooth and bounded from below by a positive number is enough to impose the vanishing at infinity of functions in $\Dom{q_1^\dagger}$ and so to guarantee the self-adjointness of $q_1^\dagger$, however, it is not: the Kempf-Mangano-Mann formulation provides a counterexample in which $q_1^\dagger$ is not self-adjoint and $f(u)=1+\beta u^2$, where $\beta$ is a positive real parameter.

\medskip

Once again, by the same reasoning, we can carry out the same computations in $\mathcal{H}_2$ and identify the domain of the bi-adjoint operator $q_2^{\dagger\dagger}$:
\begin{align} \label{biadj_q2_dom}
\begin{aligned}
     \Dom{q_2^{\dagger\dagger}}:= & \biggl\{ \phi(v) \in L^2\left(\mathbb{R}, dv \right)\ | \  2 f(v) \partial_{v} \phi (v)+\phi(v) \partial_{v} f(v) \in  L^2\left(\mathbb{R}, dv\right),
     &\lim_{v \to \pm \infty}  \sqrt{f(v)}\phi(v)=0\biggr\}.
\end{aligned}
\end{align}
We start from $\Dom{q_2}=C^{\infty}_c(\R)\subset L^2(\R,dv)$. The condition for $\psi$ to be in domain of $q_2^{\dagger}$ is that there exists $\chi\in L^2(\R,dv)$ such that:
\begin{align}
    i \hbar \int_{\mathbb{R}} dv \, \psi^{*}(v) \left(\frac{1}{2}\frac{df}{dv}\varphi(v)+f(v)\frac{d\psi}{dv}\right)=\int_{\mathbb{R}} dv\,\chi^{*}(v) \varphi(v), \, \forall \varphi \in \Dom{q_2}.
\label{Xeqn68-A.20}
\end{align}
By definition of weak derivative, we notice that the condition $\psi\in\Dom{q_2^\dagger}$ can be written as:
\begin{align}
    \int_\R|\psi(v)|^2dv<+\infty,\qquad \int_\R \left|\sqrt{f(v)}\frac{d}{dv}\left(\sqrt{f(v)}\psi(v)\right)\right|^2dv<+\infty.
\label{Xeqn69-A.21}
\end{align}
As we stated below, since the deformed measure is dominated by the Lebesgue one, we have $L^2(\R,dv)\subset L^2\left(\R,\frac{dv}{f(v)}\right)$, then:
\begin{align}
    \int_\R \left|\frac{d}{dv}\left(\sqrt{f(v)}\psi(v)\right)\right|^2dv=\int_\R\left|\sqrt{f(v)}\frac{d}{dv}\left(\sqrt{f(v)}\psi(v)\right)\right|^2\frac{dv}{f(v)}\leq \frac{1}{c}\int_\R \left|\sqrt{f(v)}\frac{d}{dv}\left(\sqrt{f(v)}\psi(v)\right)\right|^2dv<+\infty.
\label{Xeqn70-A.22}
\end{align}
Hence, the function $G(v)$ defined below is a uniformly continuous function on $\R$, and $G(v)=\sqrt{f(v)}\psi(v)-\sqrt{f(a)}\psi(a)$ almost-everywhere;
\begin{align}
    G(v)=\int_a^v\frac{d}{ds}\left(\sqrt{f(s)}\psi(s)\right)ds.
\label{Xeqn71-A.23}
\end{align}
Hence, we can consider $\sqrt{f}\psi$ uniformly continuous, and by a similar argument as above, it is bounded. In this case in enough considering the scalar product
\begin{align}
    |\bra\psi^*,\sqrt{f}(\sqrt{f}\psi)'\ket_{L^2([0,a],dv)}|\leq ||\psi||_{L^2([0,a],dv)}||\sqrt{f}(\sqrt{f}\psi)'||_{L^2([0,a],dv)}=M<+\infty,
\label{Xeqn72-A.24}
\end{align}
that reads simply as:
\begin{align}
    \int_0^a\psi(v)\sqrt{f(v)}\frac{d}{dv}\left(\sqrt{f(v)}\psi(v)\right)dv=\frac{1}{2}\left(f(a)\psi(a)^2-f(0)\psi(0)^2\right),
\label{Xeqn73-A.25}
\end{align}
concluding that $\sqrt{f(v)}\psi(v)$ in bounded. Thus, considering the condition to be in the domain of $q_2^{\dagger\dagger}$:
\begin{align}
    \begin{split}
        &\phi \in \Dom{q_2^{\dagger\dagger}} \subset L^2\left(\mathbb{R}, {dv}\right)  \iff \exists  \chi \in L^2\left(\mathbb{R}, dv\right): \\
  & i \hbar \int_{\mathbb{R}} dv \, \phi^{*}(v) \sqrt{f(v)}\frac{d}{dv}\left(\sqrt{f(v)}\psi(v)\right)=\int_{\mathbb{R}} dv\chi^{*}(v) \psi(v), \, \forall \psi \in \Dom{q_2^\dagger},
    \end{split}
\label{Xeqn74-A.26}
\end{align}
and performing integration by part
\begin{align}
  \phi(v)^{*}f(v)\psi(v)\bigg|^{+\infty}_{-\infty}- \int_{\mathbb{R}} dv \, \psi(v) \sqrt{f(v)}\frac{d}{dv}\left(\sqrt{f(v)}\phi(v)^*\right)=\frac{1}{i \hbar}\int_{\mathbb{R}} dv\chi^{*}(v) \psi(v),
\label{Xeqn75-A.27}
\end{align}
we see that the sufficient condition to have a vanishing boundary term is $\lim_{v\to\pm\infty}\sqrt{f(v)}\phi(v)=0$.

\medskip

In view of the equality $q_i^{\dagger\dagger}=\overline{q}_i$, the domains \eqref{adj_q1_dom} and \eqref{adj_q2_dom} are therefore the closure of the domain of the operator $q_1,q_2$ defined in their respective Hilbert space.
It is worth stressing that the boundary conditions we have obtained in the two settings represent a real additive request to the behavior of the involved functions only for those GUP theories where the position operators $q_i$ are \emph{not} essentially self-adjoint.
It is indeed straightforward to realize that, whenever, within our theory, the operators $q_i$ are essentially self-adjoint, we necessarily have that $q_i^{\dagger}=q_i^{\dagger\dagger}$. This means that the boundary conditions at infinity, in this case, \emph{must} be redundant, i.e., they are automatically satisfied in view of the other conditions expressed in the definition of the domains.
Ultimately, the function $f$ controls and determines this aspect of the theory.
In principle, a further, careful analysis could identify those requirements for the $f$ function to have (or not to have) redundant conditions, but, in virtue of the functional aspects we have discussed, we already know that all the information we need is actually stored in the integral appearing in \eqref{sol_def_ind_eqs}.

\section{Domains of the commutator}
\label{App:comm}

In this appendix, we verify that the second requirement of Def.~\ref{def:gup-th} is satisfied in both functional settings.
We will consider position operators to be the self-adjoint extensions $q_i^{\lambda}$. Note that if $q_i$ is essentially self-adjoint, then $q_i^{\lambda}=q^{\dagger}$, for all $\lambda$. First we want  to prove that $\Dom{[q_1^{\lambda},p_1]}\subset\Dom{f(p_1)}$, in the Hilbert space $\mathcal{H}_1$.
Here we have the following:
\begin{align}
    &\Dom{f(p)}=\{\psi(u) \in \mathcal{H}_1 | \, f(u) \,\psi(u) \in \mathcal{H}_1\}, \\
    &\Dom{[q_1^{\lambda},p_1]}=\Dom{q_1^{\lambda}p_1} \cap \Dom{p_1q_1^{\lambda}}.
\end{align}
Considering the domains in Eqs.~{\eqref{sa_ext_q}}, \eqref{mom_dom_1}, and \eqref{mom_dom_2}, as well as the domain \eqref{biadj_q1_dom}, we can write:
\begin{align}
    &\Dom{q_1^{\lambda}p_1}=\{\psi \in \mathcal{H}_1 | \, u \,\psi, f(\psi + u \psi') \in \mathcal{H}_1\ + \text{bound. terms}\}, \\
    &\Dom{p_1q_1^{\lambda}}=\{ \psi \in \mathcal{H}_1 | \,f \,\psi', u f \,\psi' \in \mathcal{H}_1 +\text{bound. terms}\}.
\end{align}
This allows us to write the domain of the commutator explicitly as:
\begin{align}
    \Dom{[q_1^\lambda,p_1]}=\{\psi \in \mathcal{H}_1| u \psi, f \psi ', u f \psi', f(\psi +u f \psi')\in \mathcal{H}_1 + \text{bound. terms}\}
\label{Xeqn76-B.5}
\end{align}

Now, since $\mathcal{H}_1:= L^2\left(\mathbb{R}, \frac{du}{f(u)}\right)$ is a vector space, $f \psi= f(\psi+u f \psi')-u f \psi'$ is in $\mathcal{H}_1$.
Hence, in the end, we have:
\begin{align}
    \Dom{[q_1^\lambda,p_1]}=\{\psi \in \mathcal{H}_1| u \psi, f \psi ', u f \psi', f\psi\in \mathcal{H}_1 + \text{bound. terms}\},
\label{Xeqn77-B.6}
\end{align}
from which the domains' inclusion relation to prove follows.

In the same way, we can verify the domains relation stated in Def.~\ref{def:gup-th}, that is $\Dom{[q_2^{\lambda},p_2]}\subset\Dom{f(p_2)}$, in $\mathcal{H}_2$.
Here we have:
\begin{align}
    \Dom{f(p_2)}&=\{\phi(v) \in \mathcal{H}_2 | \, f(v) \,\phi(v) \in \mathcal{H}_2\}, \\[1em]
        \Dom{[q_2^{\lambda},p_2]}&=\Dom{q_2^{\lambda}p_2}\cap\Dom{p_2q_2^{\lambda}} \nonumber \\
     & =\{ \phi \in \mathcal{H}_2 | \, v \phi, \, 2 f \phi' +f' \phi, \, 2 vf \phi' +v f' \phi, \\
      &  \qquad 2f \phi + 2 f v \phi'+v f' \phi \in \mathcal{H}_2  + \text{bound. terms}\} \nonumber
\end{align}
By the same argument above, we can write the domain of the commutator as:
\begin{align}
\Dom{[q,p]}=\{\phi \in \mathcal{H}_2 | \, v \phi, 2 f \phi' +f' \phi, 2 vf \phi' +v f' \phi, f \phi \in \mathcal{H}_2  + \text{bound. terms}\}
\label{Xeqn78-B.9}
\end{align}
from which the inclusion relation in this setting can be inferred as well.

\bibliographystyle{elsarticle-num} 
\bibliography{bibliography}

\end{document}